\documentclass[3p,times,twocolumn]{elsarticle}

\usepackage{ecrc}

\usepackage[figuresright]{rotating}
\usepackage[T1]{fontenc} 
\usepackage{comment} 
\usepackage{todonotes}
\usepackage{amsmath}
\usepackage{amssymb}
\usepackage{bm}
\usepackage{natbib}
\usepackage{bbold}
\usepackage{epsfig}
\usepackage{hyperref}
\usepackage{graphicx}
\usepackage{csquotes}
\usepackage{mathtools}
\usepackage{slashed}
\usepackage{color,colordvi}
\usepackage{soul}
\usepackage{xcolor}
\usepackage{multirow}
\usepackage{tabularx,ragged2e,booktabs}
\usepackage{comment}
\usepackage{slashed}
\usepackage{float}
\usepackage{subcaption}
\usepackage{amsthm}
\usepackage{mathtools}
\usepackage{physics}
\usepackage{xcolor}
\usepackage{graphicx}
\usepackage[T1]{fontenc}
\usepackage{lipsum}
\usepackage{csquotes}
\usepackage[normalem]{ulem}
\renewcommand{\(}{\left(}
\renewcommand{\)}{\right)}
\renewcommand{\{}{\left\lbrace}
\renewcommand{\}}{\right\rbrace}

\newcommand{\del}{\partial}

\newcommand{\ord}[1]{\mathcal{O}\({#1}\)}
\newcommand{\msbar}{\overline{\text{MS}}}
\newcommand{\as}{\alpha_\mathrm{s}}

\newcommand{\ordas}[1]{\mathcal{O}\left(\alpha_s^{#1}\right)}
\newcommand{\GeV}{\,\mathrm{GeV}}

\newcommand{\MeV}{\,\mathrm{MeV}}

\newcommand{\hs}{\hspace{.4mm}}

\newcommand{\bs}{\hspace{1cm}}

\volume{00}

\firstpage{1}

\journalname{Nuclear and Particle Physics Proceedings}

\runauth{}

\jid{nppp}

\jnltitlelogo{Nuclear and Particle Physics Proceedings}

\usepackage{amssymb}

\usepackage[figuresright]{rotating}

\begin{document}

\begin{frontmatter}



\dochead{}

\title{Renormalization Group Improvement and QCD Sum Rules}


\author{M. S. A. Alam Khan}
\address{Centre for High Energy Physics, Indian Institute of Science, Bangalore 560 012, India}
\ead{alam.khan1909@gmail.com}{}
\begin{abstract}
We summarize the results obtained for the quark masses (u,d,s,c, and b) in Refs.~\cite{AlamKhan:2023ili,AlamKhan:2023kgs} and strong coupling ($\as$) using renormalization group (RG) improvement of the theoretical expressions and experimental inputs that enter in the QCD sum rules. We obtain $m_{u}(2\GeV)=2.00_{-0.40}^{+0.33}\hs\MeV$, $m_{d}(2\GeV)=4.21_{-0.45}^{+0.48}\hs\MeV$, and $m_{s}(2\GeV)=104.34_{-4.24}^{+4.32}\hs\MeV$ using Borel Laplace sum rules for the divergence of the axial vector currents. The relativistic sum rules for the moments of the heavy quark currents lead to the determination of $\as(M_Z)=0.1171(7)$, $\overline{m}_{c}=1281.1(3.8)\MeV$ and $\overline{m}_{b}=4174.3(9.5)\MeV$. 
\end{abstract}


\end{frontmatter}


\section{Introduction}
The QCD sum rules~\cite{Shifman:1978bx,Shifman:1978by} play a key role in the determination of the standard model (SM) parameters. These parameters include coupling constants, masses of the particles, and higher dimensional condensates. Quark masses and coupling constants are the parameters of the QCD Lagrangian while the condensates terms appear when current correlators are expanded in terms of the fields relevant for a process using operator product expansion (OPE)~\cite{Wilson:1969zs}. In the case of perturbative QCD (pQCD), the strong coupling constant is large in the low-energy regions. The determinations of these parameters suffer from convergence and renormalization scale dependence issues. These issues, to some extent, can be controlled using prescriptions such as renormalization group summed perturbation theory (RGSPT) which uses the renormalization group equation (RGE) to sum the large running logarithms~($\log(\mu^2/Q^2)$) present in the perturbative series such that $\as(\mu^2) \log(\mu^2/Q^2)\sim \ord{1}$. These logarithms are often resummed to all orders by setting $\mu^2=Q^2$, and the evolution of the operators is performed numerically using RGE. Even in this case, it is not guaranteed that the scale dependence, which is calculated by setting $\mu^2=\xi Q^2$ and varying $\xi \in\{1/2,2\}$, will be minimum~\cite{Ananthanarayan:2020umo,Ananthanarayan:2022ufx,Ananthanarayan:2016kll}. Apart from this, it has been found that RGSPT can be used not only for RG improvement but can also sum large kinematical $\pi^2-$terms arising due to the analytic continuation of perturbative series from spacelike regions to timelike regions~\cite{AlamKhan:2023dms}. These terms are also found to be useful in estimating the higher order relation between pole and $\msbar$ quark masses in Ref.~\cite{Kataev:2019zfx} and other processes in renormalon motivated method in Ref.~\cite{Broadhurst:2000yc}. The analytic continuation is an important ingredient for Borel-Laplace and Laplace type sum rules~\cite{Narison:2023ntg,Narison:2023srj,Albuquerque:2023bex,Narison:2021xhc,Narison:2014vka}.\par
In section~\ref{sec:rgspt}, we discuss the RG improvement of the perturbative series in the RGSPT prescription. In section~\eqref{sec:QCD_rule1}, discuss the role of RG improvement in the determination of the light quark masses using the Borel-Laplace sum rule. In section~\eqref{sec:QCD_rule2}, determination of $\as$, $m_c$ and $m_b$ from the relativistic sum rules is discussed. In section~\eqref{sec:conclusion}, we give a short conclusion about the role of RG improvement in the QCD sum rule determinations. 
\section{RG improvement using RGSPT} \label{sec:rgspt}
In FOPT prescription, a finite order perturbative series $\mathcal{S}(Q^2,\mu^2)$ in pQCD can be written as:
\begin{equation}
	\mathcal{S}(Q^2,\mu^2)\equiv \sum_{i=0,j=0}^{j\le i} T_{i,j} x^i L^j \,,
	\label{eq:Pseris}
\end{equation}
where $x=\as(\mu)/\pi$ and $L=\log(\mu^2/Q^2)$.
 The RG evolution of the perturbative series in Eq.~\eqref{eq:Pseris} is obtained using its anomalous dimension, $\gamma_S(x)$, by solving:
\begin{align}
	\mu^2 \frac{d}{d\mu^2} \mathcal{S}(Q^2)&=\gamma_S(x) \hs\hs \mathcal{S}(Q^2)\,,
	\label{eq:RGE}\\
		\mu^2 \frac{d}{d\mu^2} x(\mu)&=\beta(x)\,,
\end{align}
where anomalous dimension $\gamma_S(x)$ and $\beta(x)$ is given by:
\begin{equation}\label{eq:anom_S}
	\begin{aligned}
	\gamma_S(x)&=\sum_{i=0}\gamma_i x^{i+1}\,,\\
	\beta(x)&=-\sum_{i=0}\beta_i x^{i+2}\,.	
	\end{aligned}
\end{equation}
 In RGSPT, perturbative series in Eq.~\eqref{eq:Pseris} is arranged as follows:
 \begin{equation}
 	\mathcal{S}^\Sigma(Q^2)= \sum_{i=0} x^i S_i (x\hs L)\,,
 	\label{eq:summed_ser}
 \end{equation}
 and the goal is to obtain a closed-form expression for coefficients:
 \begin{align}
 	S_i (z)=\sum_{j=0}^{\infty} T_{i+j,j} z^j\,,
 	\label{eq:SRcoef}
 \end{align}
where $z\equiv x \hs L\hs$. The coefficients $S_i (z)$ are function of one variable where $z\sim\mathcal{O}(1)$. The closed-form solution for them is obtained using RGE. \par
 The RGE in Eq.~\eqref{eq:RGE} results in a set of coupled differential equations for $S_i(z)$, which in compact form can be written as:
\begin{equation}
  \left(\sum_{i=0}^{n} \frac{\beta_{i}}{z^{n-i-1}}\frac{d}{d\hs z}\left(z^{n-i} S_{n-i}(z)\right)+\gamma_i S_{n-i}(z)\right)-S'_n(z)=0\,,
\end{equation}
The first three coefficients can be obtained by solving the above differential equation and are given by:
 \begin{align}
     S_0(z)=&T_{0,0} w^{-\tilde{\gamma }_0}\,,\nonumber\\ S_1(z)=&T_{1,0} w^{-\tilde{\gamma }_0-1}+T_{0,0} w^{-\tilde{\gamma }_0-1} \Big[(1-w) \tilde{\gamma }_1\nonumber\\ &\bs+\tilde{\beta }_1 \tilde{\gamma }_0 (w-\log (w)-1)\Big]\nonumber\\
  S_2(z)=&T_{2,0} w^{-\tilde{\gamma }_0-2}-T_{1,0} w^{-\tilde{\gamma }_0-2} \Big[(w-1) \tilde{\gamma }_1\nonumber\\&+\tilde{\beta }_1 \left(\tilde{\gamma }_0 (-w+\log (w)+1)+\log (w)\right)\Big]\nonumber\\&+\frac{1}{2} T_{0,0} w^{-\tilde{\gamma }_0-2} \Big\lbrace-\tilde{\beta }_1 \tilde{\gamma }_1 \Big[1-w^2+2 \log (w)\nonumber\\&+2 (w-1) \tilde{\gamma }_0 (w-\log (w)-1)\Big]+(w-1)\nonumber\\&\times \Big[(w-1) \tilde{\beta }_2 \tilde{\gamma }_0+(w-1) \tilde{\gamma }_1^2-(w+1) \tilde{\gamma }_2\Big]\nonumber\\&+\tilde{\beta }_1^2 \tilde{\gamma }_0\big(\tilde{\gamma }_0-1\big) (w-\log (w)-1)^{2} \Big\rbrace\,,
 \end{align}
 where $w\equiv 1-\beta_0 \hs z$, for anomalous dimension and higher order beta function coefficients, we have used $\tilde{X}\equiv X/\beta_0$. The most general term of RGSPT is given by:
 \begin{equation}
     \Omega_{n,a}(Q^2)\equiv\frac{\log^n(w)}{w^a}=\frac{\log^n(1-\beta_{0}\hs x(\mu)\log(\mu^2/Q^2))}{(1-\beta_{0}\hs x(\mu)\log(\mu^2/Q^2))^a}\,,
     \label{eq:coef_rgspt}
 \end{equation}
 where $n$ is a positive integer and $a\propto \gamma_0/\beta_{0}$ appearing in Eq.~\eqref{eq:anom_S}. The imaginary part of $\mathcal{S}(Q^2,\mu^2)$ is obtained as:
\begin{align}
\text{Im}(\mathcal{S}(s,\mu^2))=&\frac{1}{2\pi i}(\mathcal{S}(s+I \epsilon,\mu^2)-\mathcal{S}(s-I \epsilon,\mu^2)) \nonumber\\=&\frac{1}{2\pi i}\int_{s-I \epsilon}^{s+I \epsilon}d\hs q^2\frac{d}{d \hs q^2}\hs \mathcal{S}(q^2,\mu^2)\\=&\frac{1}{2\pi i}\oint_{|q^2|=s}d\hs q^2\frac{d}{d \hs q^2}\hs \mathcal{S}(q^2,\mu^2)
\end{align}
where contour integral is evaluated by avoiding the cut for $s>0$ present in $\mathcal{S}(q^2,\mu^2)$ due to logarithms of $\log(\frac{\mu^2}{-q^2})$. It should be noted that we have used variables $Q^2=-q^2>0$ for the spacelike and $s=q^2>0$ for timelike regions.\par
For FOPT, the imaginary part is obtained by taking discontinuity of $\log(\mu^2/Q^2)=\log(\mu^2/|Q|^2)\pm i \hs \pi$ results in large ``$i\hs \pi$" corrections. In the case of RGSPT, the imaginary part for the terms in Eq.~\eqref{eq:coef_rgspt} is obtained as:
\begin{align}
\frac{1}{2\pi i} &\oint_{|q^2|=s} \frac{dq^2}{q^2}\frac{\log^m(1-u_1 \log\left(\frac{\mu^2}{-q^2}\right))}{\left(1-u_1 \log\left(\frac{\mu^2}{-q^2}\right)\right)^n}\nonumber\\&=\lim_{\delta\rightarrow0}\partial_{\delta}^m 
\begin{cases}\frac{\tan ^{-1}\left(\frac{\pi u_1}{1- u_1\hs L_s}\right)}{\pi  u_1}\,, & n=1 \\
\frac{w_s^{-\frac{1}{2} (n-\delta -1)} \sin \left((n-\delta -1) \tan ^{-1}\left(\frac{\pi u_1}{1-u_1\hs  L_s }\right)\right)}{\pi u_1 (n-\delta -1)}\,,& n\neq 1
\end{cases}
\label{eq:master_eq}
\end{align}
where, $w_s=\left(1- u_1 \hs L_s  \right)^2+\pi^2 u_1^2$. We can see that all the kinematical $\pi^2$-terms are summed in $w_s$ and $\tan^{-1}(\frac{\pi u_1}{1-u_1 \hs L_s}) $. The large logarithms are also under control as they are always accompanied by $\as$, which is one of the important features of the RGSPT. Hence, both RG improvement as well as all-order summation of $\pi^2-$terms is naturally achieved in the RGSPT. For more details on analytic continuation using FOPT and RGSPT, we refer to Ref.~\cite{AlamKhan:2023dms}.\par 
One important point to note here is that results from different prescriptions, such as RGSPT and FOPT, are not the same when $\mu^2=Q^2$ is set after operations like analytic continuation or Borel transformation are performed. This is one of the motivations behind our determination of the light quark masses in Ref.~\cite{AlamKhan:2023ili}.
  \section{Borel-Laplace Sum rule and Light quark mass determinations}\label{sec:QCD_rule1}
  There are various versions of QCD sum rules are used in the literature. In this section, we discuss the application of RGSPT in the Borel-Laplace sum rules in the light quark mass determination. 
  \subsection{Basic definitions}
	The current correlator for the divergence of the axial currents is defined as:
		\begin{align}
		\Psi_{5}(q^2)\equiv i \int d^4x\hs e^{i q x} \langle 0|\mathcal{T}\{j_{5}(x) j_{5}^{\dagger}(0)\} |0\rangle\,,
  \label{eq:def_corr}
	\end{align}
	where $j_{5}$ is given by:
	\begin{align}
j_{5}=\partial^{\mu}\left(\overline{q}_1\gamma_{\mu} \gamma_5q_2\right)&=i \left(m_1+ m_2\right)\left(\overline{q}_1\gamma_5 q_2\right)\nonumber\\&=i \left(m_1+ m_2\right)j_{0}\,.
\label{eq:der_J}
	\end{align}
	 Using Eq.~\eqref{eq:der_J}, the correlation function in Eq.~\eqref{eq:def_corr}, is related to the pseudoscalar correlation function ($\Pi_P(q^2)$), by relation:
  \begin{align}
      \Psi_{5}(q^2)=\left(m_1+ m_2\right)^2\Pi_P(q^2)\,,
      \label{eq:Psi2Pi}
  \end{align}
  and
  \begin{align}
		\Pi_{P}(q^2)= i \int d^4x\hs e^{i q x} \langle 0|\mathcal{T}\{j_{0}(x) j_{0}^{\dagger}(0)\} |0\rangle\,.\nonumber
 	\end{align}
  Using OPE, a theoretical expression for $\Psi_5(q^2)$ is calculated in the deep Euclidean spacelike regions in the limit $m_q^2 \ll q^2$, and the resulting expansion can be arranged as expansion in $1/(q^2)$. At low energies $\sim 1\GeV^2$, instanton effects become relevant, and their contribution is not captured by OPE expansion and, therefore, are added to OPE contributions.\par
  	The Borel-Laplace sum rules are based on the double-subtracted dispersion relation for the correlation function. Therefore, it involves the double derivative of $\Psi_5(q^2)$ and the dispersion relation is given by:
	\begin{align}
		\Psi''_{5}(q^2)=\frac{d^2}{d (q^2)^2}\Psi_{5}(q^2)=\frac{2}{\pi}\int_{0}^{\infty}ds\frac{\text{Im}\Psi_{5}(s)}{(s-q^2-i\epsilon)^{3}}\,.
		\label{eq:der2Psi}
	\end{align}
The Borel transformation~\cite{Jamin:1994vr}, with parameter ``u",  is obtained using the Borel operator, $\hat{\mathcal{B}_u}$, defined as:
\begin{align}
	\hat{\mathcal{B}}_u\equiv\lim _{Q^{2}, n \rightarrow \infty \atop Q^{2} / n=u}\frac{(-Q^{2})^n}{\Gamma[n]}\partial_{Q^{2}}^n
 \label{eq:BO}\,.
\end{align}
  Borel parameter $u$ has the dimension of $\GeV^2$ and the Borel transform of Eq.~\eqref{eq:der2Psi} is obtained as:
\begin{align}
	\Psi''_{5}(u)&\equiv\hat{\mathcal{B}}_u\left[\Psi''_{5}(q^2)\right]=\frac{1}{u^3}\hat{\mathcal{B}}_u\left[\Psi_{5}(q^2)\right](u)\nonumber\\&=\frac{1}{\pi u^3}\int_{0}^{\infty} ds\hs e^{-s/u} \hs \text{Im}\Psi_{5}(s)\nonumber\\&=\frac{1}{u^3}\int_{0}^{\infty}ds \hs\hs e^{-s/u} \rho_{5}(s)\,,
	\label{eq:bs_rule}
\end{align}
where the spectral density is given by:
\begin{equation}
    \rho_{5}(s)=\frac{1}{\pi}\lim_{\epsilon\rightarrow 0}\left[\text{Im}\Psi_{5}(s+i\hs\epsilon)\right]\,.
    \label{eq:spectral_density}
\end{equation}
It should be noted that the value of the $u\gg\Lambda_{QCD}^2$ in $\Psi_{5}''(u)$ is chosen such that higher order terms of the OPE expansion remain suppressed in the Borel transformed OPE expansion. \par
The spectral density in the RHS of Eq.~\eqref{eq:bs_rule} Borel-Laplace sum rules can be decomposed in the following form:
\begin{equation}
    \rho_5(s)=\theta(s_0-s)\rho_5^\text{had.}+\theta(s-s_0)\rho_5^{\text{OPE}}\,,
    \label{eq:spectral_decomposition}
\end{equation}
where scale $s_0$ separates the two contributions from hadronic states and continuum. The Borel sum rule in Eq.~\eqref{eq:bs_rule} can be written as:
\begin{align}
		\Psi''_{5}(u)&=\frac{1}{u^3}\int_{0}^{s_0}ds \hs\hs e^{-s/u} \rho^{had}_{5}(s)\nonumber\\&\hs\bs+\frac{1}{u^3}\int_{s_0}^{\infty}ds \hs\hs e^{-s/u} \rho^{OPE}_{5}(s)\,.
		\label{eq:bs_final}
\end{align}
 which is used in this article for the light quark mass determination. \par 
For clarification, various inputs used in Eq.~\eqref{eq:bs_final} are as follows:
\begin{enumerate}
\item \label{item:1}The $\Psi''_{5}(u)$ is obtained from the Borel transformation of $\Psi''(q^2)$, which involves OPE corrections and addition to the instanton contributions. 
\item \label{item:2} The hadronic spectral density $\rho_{5}^{\text{had}}(s)$ is obtained by the parametrization of the experimental information on the hadrons appearing in the strange and non-strange channels. 
\item \label{item:3}$\rho_5^{\text{OPE}}(s)$ in the RHS of Eq.~\eqref{eq:bs_final} is obtained from the discontinuity of the theoretical expression of the $\Psi_5(q^2)$. 
\end{enumerate}
It should be noted that our main focus is the RG improvement for the theoretical quantities relevant for point~\eqref{item:1} and point~\eqref{item:3} and its impact on the light quark mass determination. 
\subsubsection{RG improvement of pQCD inputs}
\begin{figure}[ht]
    \centering
  \includegraphics[width=\linewidth]{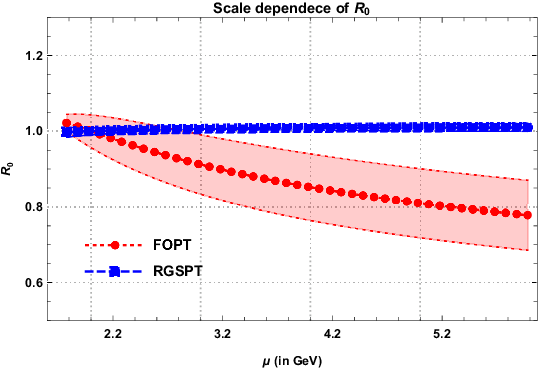}
		\includegraphics[width=\linewidth]{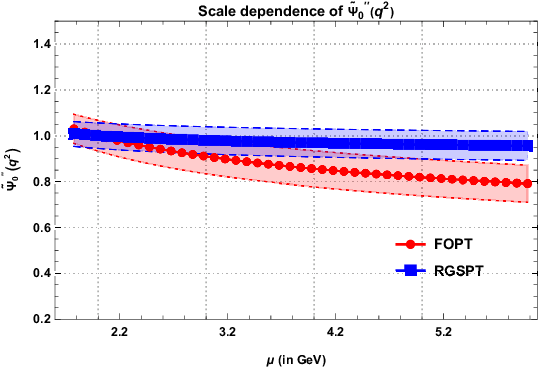}
    \caption{   \label{fig:scdepR0Pd2}Renormalization scale dependence of $R_0(s)$ and $\overline{\Psi}''_0(q^2)$ normalized to unity at $2\hs\GeV$ in RGSPT and FOPT. The bands represent the truncation uncertainty.}
\end{figure}

The spectral density from the OPE expansion of the polarization function, its second derivative, and the Borel transform of the second derivative can be written as
    \begin{align}
	\rho^{OPE}_{5}(s)&= s \hs \mathcal{R}_0(s)+\mathcal{R}_2(s)+\frac{1}{s}\mathcal{R}_4(s)+\cdots\,,\label{eq:eq1}\\
 \Psi''_{5}(Q^2)&=\frac{1} {Q^2}\sum_{i=0}\frac{\tilde{\Psi}''_i(Q^2)}{(Q^2)^i}\,\label{eq:eq2}\\
      \Psi''_{5}(u)&=\frac{1} {u}\sum_{i=0}\frac{\tilde{\Psi}''_i(u)}{u^i}\label{eq:eq3}\,.
\end{align}
 The leading OPE contributions in Eq.~\eqref{eq:eq1} and Eq.~\eqref{eq:eq2} are known to $\ordas{4}$. For more details on the various contributions, we refer to Ref.~\cite{AlamKhan:2023ili}. For these quantities, we define:  
\begin{align}  
R_0(s)&\equiv\frac{8\pi^2 }{3 (m_s(2\GeV))^2}\mathcal{R}_0(s)\,,\\  \overline{\Psi}''_0(Q^2)& \equiv\frac{8\pi^2 }{3 (m_s(2\GeV))^2}\tilde{\Psi}''_0(Q^2)\,.
\end{align}
 Using $\as(2\GeV)=0.2945$ and $m_s(2\GeV)=93.4\MeV$ at $\mu=2\GeV$ and setting $m_u=0$, the above quantities in FOPT and RGSPT at $Q^2=2\GeV^2$ has contributions from different orders:
\begin{align}
    R_0^{\text{FOPT}}&=1.0000+0.6612+ 0.4909+ 0.2912+ 0.1105\,,\nonumber\\
    R_0^{\text{RGSPT}}&=1.0038+0.4175+0.1760+ 0.0581 -0.0152\,,\nonumber\\
\overline{\Psi}^{'',\text{FOPT}}_0&=1.0000+0.4737+ 0.2837+ 0.1917+ 0.1405\,,\nonumber\\
  \overline{\Psi}^{'',\text{RGSPT}}_0&=1.1508+ 0.5280+ 0.2621+ 0.1670+ 0.1244\nonumber\,.
\end{align}
The scale dependence of $R(s)$ and $\overline{\Psi}$ is shown in Fig.~\eqref{fig:scdepR0Pd2}. We see that RGSPT gives a better convergence than FOPT for $\rho_5^{\text{OPE}}$.\par
Next, we calculate the Borel transform for FOPT and RGSPT. 
In FOPT, Borel transform is given by:
        \begin{align}
    \hat{\mathcal{B}}_{u}\Big[\frac{1}{(Q^2)^\alpha}&\log^n\left(\frac{\mu^2}{Q^2}\right)\Big]\nonumber\\&=\frac{1}{(u)^\alpha}\sum_{k=0}^n (-1)^k \hs\hs\text{}^n C_k\log^k\left(\frac{\mu^2}{u}\right)\del_\alpha^{n-k}\frac{1}{\Gamma[\alpha]}\,.
\end{align}
In RGSPT prescription, Borel transform can only be evaluated numerically by using:
\begin{align}
\hat{\mathcal{B}}_{u}\left[\frac{1}{s^{z}}\frac{1}{w^{\alpha}}\right]&=\frac{1}{(\mu^2)^z\Gamma[\alpha]}\sum_{n=0}^{\infty}\frac{(-1)^n \Gamma[\alpha+n-1]}{\Gamma[n+1] \left(\beta_0 \hs x\right)^{n+\alpha}}\nonumber\\&\hspace{2cm}\times\tilde{\mu}(\mu^2/u,\alpha+n-1,z)\,,\nonumber\\
	\tilde{\mu}(z,b,a)&\equiv\int_{0}^{\infty}dt\frac{x^{a+t}t^{b}}{\Gamma[b+1]\Gamma[a+t+1]}
	\label{eq:identity1}
\end{align}
Now, we can demonstrate the impact of the resummation for the Borel transformation. The leading mass corrections at different dimension to $\tilde{\Psi}''_j(s)$ from RGSPT has the following form:
\begin{align}
   A_j^{\text{RGSPT}}= \frac{1}{s(1-\beta_0 x L)^{(2j+2)\gamma_0/\beta_0}}\,,
\end{align}
where $L=\log(\mu^{2}/s)$ is used here for the discussion. Its series expansion to $\ordas{4}$ in FOPT is given by:
\begin{align}
   A_j^{\text{FOPT}}= &\frac{1}{s}\bigg(1+2 \gamma _0 L (j+1) x\nonumber\\&+\gamma _0 L^2 (j+1) x^2 \left(\beta _0+2 \gamma _0(1+j)\right)\nonumber\\&+\frac{2}{3} \gamma _0 L^3 (j+1) x^3 \left(\beta _0+(1+j)\gamma _0 \right)\nonumber\\&\bs\times \left(\beta _0+2 \gamma _0(1+j)\right)\nonumber\\&+\frac{1}{6} \gamma _0 L^4 (j+1) x^4\left(\beta _0+\gamma _0(1+ j)\right)\nonumber\\&\bs\times \left(\beta _0+2 \gamma _0(1+j) \right) \left(3 \beta _0+2 \gamma _0(1+j)\right)\bigg)\nonumber\\&+\ordas{5}\,.
   \label{eq:Bexp}
\end{align}

\begin{figure}[ht]
\centering
		\includegraphics[width=\linewidth]{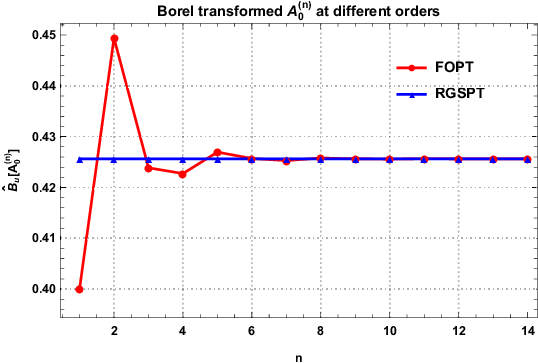}
		\includegraphics[width=\linewidth]{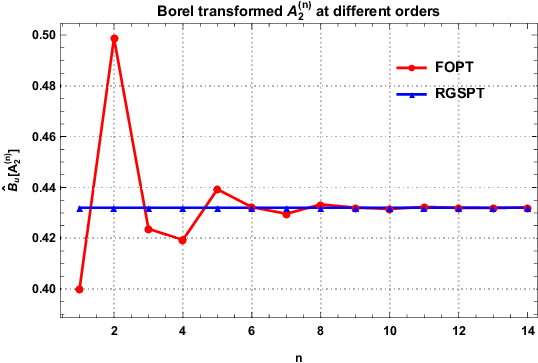}
 \caption{Borel transform of $A_0$ and $A_2$ calculated at different orders using $u=2.5\GeV^2$. }
 \label{fig:A0}
    \end{figure}
Now, we obtain the Borel transform for $A_0$ by setting $\mu^2=u=2.5\GeV^2$ that resums the logarithms in the case of FOPT. Using $x(\sqrt{2.5})=\as(\sqrt{2.5})/\pi=0.3361/\pi$, the Borel transformation of $A_0$ has the following contributions:
\begin{align}
    \hat{\mathcal{B}}_u [A_0^{\text{RGSPT}}]&=0.4256\,,\nonumber\\
     \hat{\mathcal{B}}_u [A_0^{\text{FOPT}}]&=0.4000+ 0.0494-0.0255 -0.0011\nonumber\\&\bs+ 0.0042+\cdots\nonumber\\&=0.4270\,.
\end{align}
The higher order behavior of the Borel transform of $A_0$ and $A_2$ are presented in Fig.~\eqref{fig:A0}. It is shown that the Borel transform of the leading logarithm terms in FOPT prescription needs at least $\ordas{6}$ terms to converge to the Borel transform of the leading term in RGSPT. 
\subsubsection{Hadronic inputs}
The hadronic spectral function for non-strange and strange channels is parameterized in terms of the experimental information on the hadronic states. We use the parametrization provided in Ref.~\cite{Dominguez:2018azt} for non-strange, and for the strange channel, we use the spectral function from Ref.~\cite{Dominguez:1997eu}. 
The hadronic spectral density has the following form:
\begin{align}
    \rho_{\text{NS}}&= f_\pi^2 M_\pi^4\hs \delta\left(s-M_\pi^2\right)+\rho_{3\pi} \frac{\text{BW}_1(s)+\kappa_1 \hs \text{BW}_2(s)}{1+\kappa_1}\,,\label{eq:had_NS}\\
    \rho_S(s)&=f_K^2 M_K^2\delta\left(s-M_K^2\right)+ \rho_{K\pi\pi}(s)\frac{\text{BW}_1(s)+\kappa_2 \hs \text{BW}_2(s)}{1+\kappa_2}\,.
    \label{eq:had_S}\end{align}
    where Breit-Wigner distribution normalized to unity at threshold is given by:
    \begin{align}
    \text{BW}_i(s)&=\frac{\left(M_i^2-s_{\text{th}}\right)^2+M_i^2\Gamma_i^2}{\left(s-M_i^2\right)^2+M_i^2\Gamma_i^2}\,.
\end{align}
 The values of $\kappa_1\simeq0.1$ ~\cite{Dominguez:2018azt} and $\kappa_2\simeq1$ ~\cite{Dominguez:1997eu} controls the relative importance of the resonances. Three particle contributions from the resonance regions ($\rho_{3\pi}$ and $\rho_{K\pi\pi}$ ) are calculated using chiral perturbation theory~\cite{Dominguez:1997eu,Dominguez:1986aa,Bijnens:1994ci}. Resonance contributions are parameterized using the Breit-Wigner distribution. The non-strange channel receives contributions from the $\pi(1300)$ and $\pi(1800)$ states and for strange channel $K(1460)$ and $K(1830)$ states contribute. 
\begin{figure}[ht]
\centering
\includegraphics[width=\linewidth]{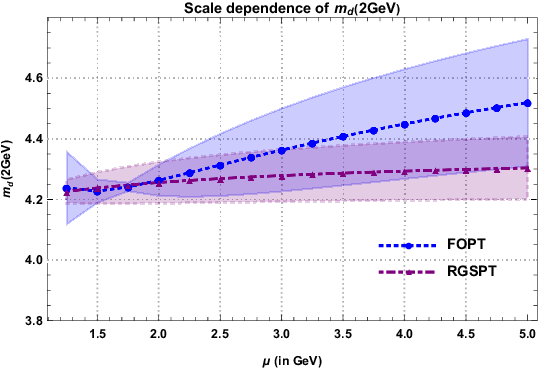}
\includegraphics[width=\linewidth]{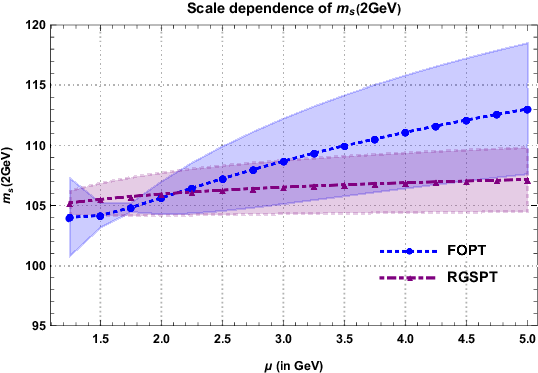}
\caption{Scale dependence in our determination of the $m_s(2\GeV)$ and $m_d(2\GeV)$. The bands represent the truncation uncertainty at different renormalization scales.}
\label{fig:scdep_msmd}
\end{figure}
\subsubsection{Light quark mass determination}
    To determine the light quark masses, we use Eqs.\eqref{eq:eq1}, \eqref{eq:eq3}, \eqref{eq:had_NS} and \eqref{eq:had_S} for strange and non-strange channels as an input for Borel-Laplace sum rule in Eq.~\eqref{eq:bs_final}. We get the most precise determination using RGSPT, where theoretical uncertainties are significantly reduced, and the values for $m_s$ and $m_d$ at $2\GeV$ are obtained as:
\begin{align}
         m_s(2\hs\GeV)&=104.34_{\left(-2.40\right)_{\text{pQCD}}
        \left(-3.50\right)_{\text{had.}}}^{\left(+2.42\right)_{\text{pQCD}}\left(+3.57\right)_{\text{had.}}}\MeV\nonumber\\&=104.34_{-4.24}^{+4.32}\hs\MeV\,,\\
         m_d(2\hs\GeV)&=4.21_{\left(-0.10\right)_{\text{pQCD}}\left(-0.43\right)_{\text{had.}}}^{\left(+0.10\right)_{\text{pQCD}}\left(+0.47\right)_{\text{had.}}}\MeV\nonumber\\&=4.21_{-0.45}^{+0.48}\hs\MeV\,,\\
        \implies  m_u(2\hs\GeV)&=2.00_{-0.40}^{+0.33}\hs\MeV\,.
\end{align}
where, we use the ratio $\epsilon_{ud}\equiv m_u/m_d=0.474_{-0.074}^{+0.056}$~\cite{ParticleDataGroup:2022pth} for the $m_u$ determination. The uncertainties from the parameters such as $\as$, $\mu$, and truncation of the perturbative series are included in pQCD uncertainties, and the rest of them are included in the hadronic uncertainties. The scale dependence in our determination is presented in Fig.~\eqref{fig:scdep_msmd}.\par
 Our determinations are in agreement with the current PDG average values~\cite{ParticleDataGroup:2022pth}:
\begin{align}
        m_s(2\hs\GeV)&=93.4_{-3.4}^{+8.6}\hs\MeV\,,\\
        m_d(2\hs\GeV)&=4.67_{-0.17}^{+0.48}\hs\MeV\,,\\
        m_u(2\hs\GeV)&=2.16_{-0.26}^{+0.49}\hs\MeV\,.
\end{align}
\section{Relativistic sum rules and \texorpdfstring{$\as$, $m_c$ and $m_b$}{} determination}\label{sec:QCD_rule2}
\subsection{Basic definitions}
The normalized total hadronic cross-section ($R_{\overline{q}q}$), defined as:
\begin{align}
	R_{\overline{q}q}\equiv\frac{3s}{4\pi\alpha^2}\sigma\left(e^+e^-\rightarrow q\overline{q}+X\right)\simeq\frac{\sigma\left(e^+e^-\rightarrow q\overline{q}+X\right)}{\sigma\left(e^+e^-\rightarrow \mu^+\mu^-\right)}\,,
 \label{eq:Rratio}
\end{align}
is one of the most important observable sensitive to the quark mass ($m_q$). The inverse moment for the vector channel ($\mathcal{M}_q^{V,n}$), are derived from $R_{\overline{q}q}$ as:
\begin{align}
	\mathcal{M}_q^{V,n}=\int\frac{ds}{s^{n+1}}R_{q\overline{q}}\,.
 \label{eq:Moments_V}
\end{align}
 Using analyticity and unitarity, the moments are related to the coefficients of the Taylor expansion for the quark-heavy correlator evaluated around $s=0$ as:
\begin{align}
	\mathcal{M}_{n}^{V,\text{th}}=\frac{12\pi^2Q^2_q}{n!}\frac{d^n}{ds^n}\Pi^V(s)\Big|_{s=0}
\end{align}
where $Q_q$ is the electric charge, $s=\sqrt{q^2}$ is the $e^+e^-$ center of mass energy, and $\Pi_V(s)$ are the current correlators of two vector currents given by:
\begin{align}
	\left(s\hs g_{\mu\nu}-q_\mu q_\nu\right)\Pi^V(s)=-i \int dx e^{i q x}\langle0\vert T\{j_\mu(x)j_\nu(0)\}\vert0 \rangle\,,\nonumber
\end{align}
where, 
\begin{align}
	j_\mu=\overline{q}(x)\gamma^\mu q(x)\,.\nonumber
\end{align}
For the pseudoscalar channel, slightly different definitions are used in Ref.~\cite{Dehnadi:2015fra}. The pseudoscalar current correlator is defined as
\begin{align}
    \Pi^{P}(s)&\equiv i\int d\hs x e^{i\hs q\hs x}\langle 0\vert T\lbrace j_P(x)\hs j_P(0)\rbrace\vert 0\rangle\,,
    \label{eq:vacP}
\end{align}
where,
\begin{align}
    j_P=2\hs i\hs m_q\hs \overline{q}(x)\gamma^5 q(x)\,,
\end{align}
and the double subtracted polarization function is obtained from Eq.~\eqref{eq:vacP} as:
\begin{align}
    P(s)=\frac{1}{s^2}\left(\Pi^{P}(s)-\Pi^{P}(0)-s\hs\left[ \frac{d}{d\hs s}\Pi^{P} (s)\right]_{s=0}\right)\,,
\end{align}
from which the moments are obtained as:
\begin{align}
    \mathcal{M}^{P,\text{th}}_n(s)=\frac{12\pi^2 Q_q^2}{n!} \frac{d^n}{d\hs s^n}P(s)\Big|_{s=0}\,.
\end{align}
Theoretical moments are calculated using the OPE and have contributions from purely perturbative ($\mathcal{M}_n^{X,\text{pert}}$) as well as non-perturbative ($\mathcal{M}_n^{X,\text{n.p}}$) origin. Therefore, we can write the theoretical moments as follows:
 \begin{align}
\mathcal{M}_{n}^{X,\text{th}}=\mathcal{M}_n^{X,\text{pert.}}+\mathcal{M}_n^{X,\text{n.p.}}\,.
\label{eq:Def_MX}
 \end{align}
 The fixed order perturbative series for $\mathcal{M}_n^{X,\text{pert.}}$ have the following form:
\begin{align}
    \mathcal{M}_n^{X,\text{pert}}=m_q^{-2n}\sum_{i=0}T_{i,j}^X x^i L^j
    \label{eq:mom_fopt}
\end{align}
where $m_q\equiv m_q(\mu)$, $x\equiv\as(\mu)/\pi$ and $L\equiv \log(\mu^2/q^2)$. \par 
The two-loop correction to $\mathcal{M}_n^{X,\text{pert}}$ are calculated in Ref.~\cite{Kallen:1955fb}, three-loops in Refs.~\cite{Chetyrkin:1995ii,Chetyrkin:1996cf,Boughezal:2006uu,Czakon:2007qi,Maier:2007yn}, the first four moments at four-loop (or $\as^{3}$)  from Refs.~\cite{Maier:2009fz,Hoang:2008qy}. Predictions for higher moments using the analytic reconstruction method can be found in Refs.~\cite{Greynat:2011zp,Greynat:2010kx} and are used in Ref.~\cite{Greynat:2012bxq} in the $m_c$ determination. Other predictions using Pad\'e approximants can be found in Ref.~\cite{Kiyo:2009gb}. A large-$\beta_0$ renormalon-based analysis for the low energy moments of the current correlators can be found in Ref.~\cite{Boito:2021wbj}.\par
 The $\mathcal{M}_n^{X,\text{n.p}}$ include the contributions from the condensate terms and has the following form:
\begin{align}
\mathcal{M}_n^{X,\text{n.p.}}=&\frac{1}{\left(2\hs m_q\right)^{4n+4}}\Big\langle\frac{\as}{\pi}G^2\Big\rangle_{\text{RGI}}\nonumber\\&\times\hs\left(T^{X,\text{n.p.}}_{0,0}+x(m_q) T^{X,\text{n.p.}}_{1,0}\right)+\ord{x^2}\,.
    \label{eq:def_cond}
\end{align}
 where, $T^{X,\text{n.p.}}_{i,0}$ are the perturbative correction as prefactors to the gluon condensate and are known to NLO~\cite{Broadhurst:1994qj}. For the RG invariant  gluon condensate, we use the following numerical value~\cite{Ioffe:2005ym}:
 \begin{align}
     \langle\frac{\as}{\pi}G^2\Big\rangle_{\text{RGI}}=0.006\pm0.012\GeV^4\,.
 \end{align}
In addition, we also need quark mass relations to one-loop~\cite{Tarrach:1980up} from the $\msbar$ scheme to the on-shell scheme given by:
\begin{align}
    m_q(\mu)=M_q \left(1-x(\mu)\left(\frac{4}{3}+\log\left(\frac{\mu^2}{M_q^2}\right)\right)\right)+\ord{x^2}\,,
\end{align}
which will be used in Eq.~\eqref{eq:def_cond} for the quark condensate terms.\par
From theoretical moments, defined in Eq.~\eqref{eq:Def_MX}, the ratio of the moments ($\mathcal{R}^{X}_n$) can be obtained as:
\begin{align}
\mathcal{R}^{X}_n\equiv\frac{\left(\mathcal{M}^{X}_n\right)^{\frac{1}{n}}}{\left(\mathcal{M}^{X}_{n+1}\right)^{\frac{1}{n+1}}}\,,
\label{eq:Def_R}
\end{align}
which are more sensitive to the $\as$ and less sensitive to the quark masses. The mass dependence arises only from the running logarithms present in the perturbative expansion. This quantity is very useful in the determination of the $\as$.
\subsection{RG improvement}
\begin{figure}[hp]	
\centering
	\includegraphics[width=\linewidth]{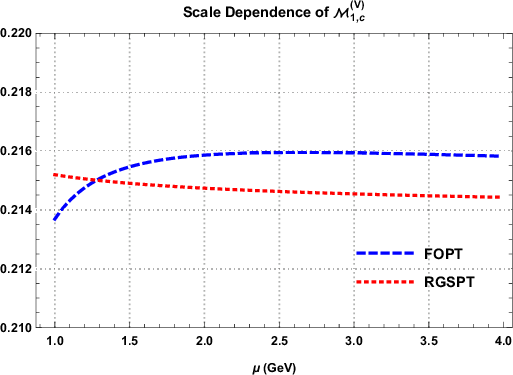}
	\includegraphics[width=\linewidth]{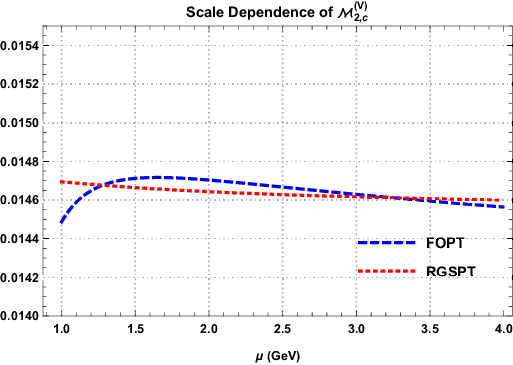}
	\includegraphics[width=\linewidth]{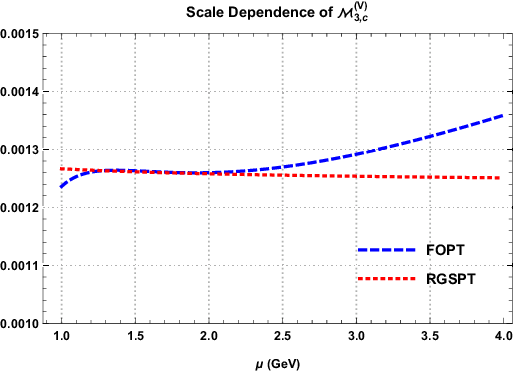}
	\includegraphics[width=\linewidth]{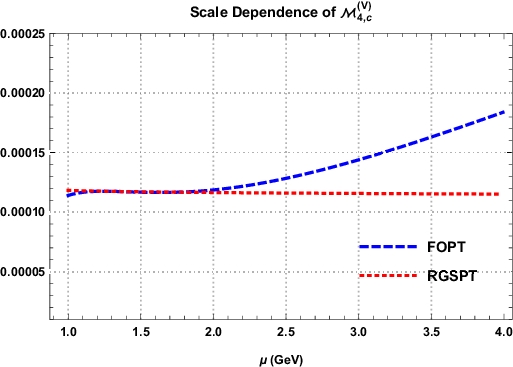}
\caption{ Renormalization scale dependence of the first four vector moments for the heavy charm quark current.}\label{fig:MomV_c}
\end{figure}

The RG improvement of moments in Eq.~\eqref{eq:mom_fopt} is obtained by rewriting moments as:
    \begin{align}
        \mathcal{M}_n^{X,\Sigma}= m_q^{-2\hs n}\sum_{i=0}x^i\hspace{.4mm}S_{i}(x \hspace{.4 mm} L)\,,
        \label{eq:ser_summed}
    \end{align}
  and following the procedure described in section~\eqref{sec:rgspt},
    we get a set of coupled differential equations for $S_{i}( x\hs L)$ that can be written in a compact form as:
    \begin{align}
        \sum _{i=0}^k \bigg[\beta _i &(\delta_{i,0}+w-1)  S_{k-i}'(w)\nonumber\\&+S_{k-i}(w) \left(-2 n \gamma_i+\beta _i (-i+k)\right)\bigg]=0\,.
        \label{eq:summed_de}
    \end{align}
After RG improved perturbative series is obtained for different $\mathcal{M}^{X}_n$, we can study their scale dependence. For the charm 
 vector moments, we take $\as^{\left(n_f=4\right)}(3\GeV)=0.2230$ and $m_c(3\GeV)=993.9\MeV$. These values are obtained from current PDG~\cite{ParticleDataGroup:2022pth} values of the quark masses and coupling constant and evolved to different scales using the REvolver package~\cite{Hoang:2021fhn}. The scale dependence of the first four moments for the vector channel for the charm case can be found in Fig.~\eqref{fig:MomV_c}, and similar behavior is obtained for the pseudoscalar channel. It should be noted that the agreement of various moments in FOPT and RGSPT prescription occurs at the $\msbar$ value of the quark masses, i.e. $\mu=m_q(\mu)=m_q(m_q)$. At this particular scale, the RGSPT expressions reduce to FOPT expressions. It is evident from these figures that the RGSPT has better control of the scale variations compared to the FOPT. 
 \subsection{\texorpdfstring{$\as$, $m_c$ and $m_b$ determinations}{}}
 For the vector moments, the third and fourth moments in the FOPT scheme are very sensitive to scale variations and contribute to a large theoretical uncertainty even though their experimental values are known more precisely~\cite{Dehnadi:2011gc,Chetyrkin:2017lif}.
  Also, the $\msbar$ definition of the quark mass for the vector channel, when used in the non-perturbative gluon condensate terms, gives unreliable determinations for the strong coupling and quark masses. This problem in FOPT prescription is cured by using the on-shell mass taken as input~\cite{Chetyrkin:2017lif,Dehnadi:2011gc} in condensate terms. However, such problems are not encountered in the determinations using RGSPT prescription. 
 With these advantages in hand, we have used FOPT and RGSPT in the determinations of the $\as$, $m_c$, and $m_b$. In the $m_c$ determination, we use experimental moments for the vector channel from Refs.~\cite{Chetyrkin:2017lif,Dehnadi:2011gc} and pseudoscalar moments from Refs.~\cite{HPQCD:2008kxl,McNeile:2010ji,Maezawa:2016vgv,Petreczky:2019ozv,Petreczky:2020tky} using lattice QCD inputs. Similarly, for the $\as-$determination, we also use the ratios of the vector moments from Refs.~\cite{Boito:2020lyp,Boito:2019pqp} and pseudoscalar moments from Refs.~\cite{HPQCD:2008kxl,McNeile:2010ji,Maezawa:2016vgv,Petreczky:2019ozv,Petreczky:2020tky,Nakayama:2016atf}. For bottom quark mass determination, we use only the vector moments from Refs.~\cite{Kuhn:2007vp,Chetyrkin:2009fv,Dehnadi:2015fra}. \par
Now, we turn to the final values for the $m_c$, $m_b$, and $\as$ determination obtained using the inputs from experimental moments or lattice QCD from the above-mentioned references. Interestingly, the most precise values of these parameters are obtained using RGSPT and lattice inputs except for the bottom quark mass, for which no lattice moments are available. \par 
For the charm mass, our final determination is obtained using moments from Ref.~\cite{Petreczky:2020tky} as:
\begin{align}
    m_c(3\GeV)&=0.9962(42)\hs\GeV\,,\\
    \implies m_c(m_c)&=1.2811(38)\hs\GeV\,.
\end{align}
For the bottom quark mass, we take the most precise value obtained from experimental moments presented in Ref.~\cite{Chetyrkin:2009fv} as:
\begin{align}
    m_b(10\GeV)&=3.6311(98)\hs\GeV\,,\\
    \implies m_b(m_b)&=4.1743(95)\hs\GeV\,.
\end{align}
We have two most precise determinations for strong coupling constant from Refs.~\cite{McNeile:2010ji,Petreczky:2019ozv}. We average out these values and obtain the final determination:
\begin{align}
    \as(M_Z)&=\{0.1172(7),0.1169(7)\}\,,\\
    \implies  \as(M_Z)&=0.1171(7)\,.
\end{align}
These values are in agreement with the current PDG~\cite{ParticleDataGroup:2022pth} values which read:
\begin{align}
    \as(M_Z)&=0.1179(9)\,,\\
     m_c(m_c)&=1.27\pm0.02\hs\GeV\,,\\
     m_b(m_b)&=4.18\pm0.03\hs\GeV\,.
\end{align}
\section{Conclusion}\label{sec:conclusion}
Precise determination of QCD parameters from QCD sum rules requires higher-order information using theoretical inputs from pQCD and precise experimental or lattice QCD simulation data. The RG improvement plays a key role in improving the theoretical inputs, and we have achieved it using the RGSPT prescription. The RGSPT provides a better convergence and enhanced stability with respect to the renormalization scale variations for the Borel-Laplace and Laplace type of sum rules. While for relativistic sum rules for heavy quark currents, one only obtains stability with respect to the renormalization scale variations crucial for the higher moments. One can also avoid unstable results when quark masses in the $\msbar$ scheme are used in the non-perturbative condensate terms using FOPT prescription.    
\section*{Acknowledgment}
 Author is supported by a scholarship from the Ministry of Human Resource Development (MHRD), Govt. of India.
\bibliographystyle{elsarticle-num}



\end{document}